\documentclass[aps,prd,superscriptaddress,amsmath,groupedaddress,twocolumn]{revtex4}
\usepackage{color,hangcaption,hhline,psfrag,rotating,amssymb}
\usepackage[hang,nooneline]{subfigure}
\usepackage{dcolumn}

\newcommand{\xls}{{x_l^\mathrm{sea}}}
\newcommand{\xss}{{x_s^\mathrm{sea}}}
\newcommand{\xl}{x_l}
\newcommand{\xs}{x_s}
\newcommand{\xa}{x_a}
\newcommand{\xb}{x_b}
\newcommand{\lamqcd}{\Lambda_\mathrm{QCD}}

\begin{document}
\title{Precise determination of the lattice spacing in full lattice QCD}

\author{C. T. H. Davies}
\email[]{c.davies@physics.gla.ac.uk}
\affiliation{Department of Physics and Astronomy, University of Glasgow, Glasgow, G12 8QQ, UK}
\author{E. Follana}
\affiliation{Department of Theoretical Physics, University of Zaragoza, E-50009 Zaragoza, Spain}
\author{I. D. Kendall}
\affiliation{Department of Physics and Astronomy, University of Glasgow, Glasgow, G12 8QQ, UK}
\author{G. Peter Lepage}
\affiliation{Laboratory of Elementary-Particle Physics, Cornell University, Ithaca, New York 14853, USA}
\author{C. McNeile}
\affiliation{Department of Physics and Astronomy, University of Glasgow, Glasgow, G12 8QQ, UK}
%\email[]{Your e-mail address}
%\homepage[]{Your web page}
%\thanks{}
%\altaffiliation{}
%\affiliation{}

%Collaboration name if desired (requires use of superscriptaddress
%option in \documentclass). \noaffiliation is required (may also be
%used with the \author command).
%\collaboration can be followed by \email, \homepage, \thanks as well.
\collaboration{HPQCD collaboration}
\homepage{http://www.physics.gla.ac.uk/HPQCD}
\noaffiliation

\date{\today}

\begin{abstract}
We compare three different methods to determine the lattice spacing in 
lattice QCD and give results from calculations on the MILC ensembles 
of configurations that include the effect of $u$, $d$ and $s$ sea quarks. 
It is useful, for ensemble to ensemble comparison, to express the 
results as giving a physical value for $r_1$, a parameter from the 
heavy quark potential. Combining the three methods gives a value
for $r_1$ in the continuum limit of  0.3133(23)(3)\,fm. 
Using the MILC values for $r_0/r_1$, this corresponds to a value for 
the $r_0$ parameter of 0.4661(38)\,fm. We also discuss how to use
the $\eta_s$ for determining the lattice spacing and tuning the $s$-quark mass accurately, 
by giving values for $m_{\eta_s}$ (0.6858(40) GeV) and $f_{\eta_s}$ (0.1815(10) GeV).
\end{abstract}

% insert suggested PACS numbers in braces on next line
%\pacs{}
% insert suggested keywords - APS authors don't need to do this
%\keywords{}

%\maketitle must follow title, authors, abstract, \pacs, and \keywords
\maketitle

\section{Introduction}

Results from lattice QCD calculations are generally computed in units of the lattice spacing~$a$ used in the simulation. The lattice spacing must be computed separately and divided out in order to convert these results into physical units (GeV, fm\,\ldots), for comparison with experiment.
Any error in the lattice spacing determination
feeds into most other
quantities from lattice QCD, and, in many 
cases, it is among the dominant sources of errors. For example, in our 
determination of the decay constant of the $D_s$ meson~\cite{fds}, 1\% of the 
total error of 1.3\% comes from the 1.5\% uncertainty in the value of 
the lattice spacing. 
Reducing the error 
on the lattice spacing is then very important for increasing the precision of the 
 realistic lattice QCD calculations now possible~\cite{ourlatqcd}.

Generally the value of the lattice spacing is determined by comparing
values from the simulation, in lattice units, with 
values from experiment, in physical units. A lattice simulation, for 
example, might give a value for the pion decay constant in lattice units:
$a f_\pi^\mathrm{lat}$. Dividing by the experimental value $f_\pi^\mathrm{exp}$ in GeV gives a value for the lattice
spacing, $a= (a f_\pi^\mathrm{lat})/f_\pi^\mathrm{exp}$, in inverse~GeV. This lattice spacing can then be used to convert other simulation results
from lattice units to physical units.

Lattice spacings determined in this way are 
inherently ambiguous because lattice simulations are 
never exact. In particular the use of a nonzero lattice spacing causes 
lattice quantites, like $f_\pi^\mathrm{lat}$, to deviate from their physical
values, in this case $f_\pi^\mathrm{exp}$. Such errors differ from quantity
to quantity, and therefore so will values for the lattice spacing
that are computed from these quantities. Such differences, however, 
vanish in the continuum limit,~$a\to0$, and so do not 
affect lattice predictions that have been extrapolated to~$a=0$. 

In principle, any dimensionful quantity can be used
 to determine the lattice spacing, but some quantities are more useful 
than others. Ideally one wants quantities that are easily computed,
free of other types of simulation error, largely independent of lattice 
parameters other than the lattice spacing, and well measured in experiments.
Use of the pion decay constant, for example, is not ideal. 
This decay constant is quite sensitive
to the $u$~and $d$~quark masses, which are generally too large in current 
simulations; accurate values for the decay constant can be obtained only after 
chiral extrapolations of the simulation data to the physical quark masses. 
This greatly complicates the use of the decay constant to set the lattice 
spacing.

One physical quantity that is very easy to calculate in lattice simulations
is the $r_1$~parameter derived from the potential~$V(r)$ between
two infinite-mass quarks separated by distance~$r$. Parameter $r_1$ is defined 
implicitly by the equation $r_1^2 F(r_1) = C$ where $F(r)\equiv dV/dr$ 
and $C=1$~\cite{milc2}. (Taking $C=1.65$ gives the original such 
standard parameter, $r_0$~\cite{sommer}.) This quantity
is easily calculated, in lattice units (that is, $r_1/a$), to better than~1\%.
Unlike the pion decay constant, it is only weakly dependent upon the
quark masses. It would be an ideal choice for setting the lattice spacing
except for the fact that there is no experimental
value for the physical~$r_1$\,---\,this must be estimated instead 
from other lattice calculations.

In this paper we examine three other quantities that can be used
to determine the lattice spacing: 1)~the radial excitation 
energy in the $\Upsilon$ system ($m_{\Upsilon^{\prime}}-m_{\Upsilon}$); 
2)~the mass difference between the $D_s$ meson and one half the 
$\eta_c$ mass; and 3)~the decay constant of the fictitious $\eta_s$ particle, 
which can be related accurately to $f_K$ and $f_{\pi}$. The valence-quark masses are easily tuned in each case and each quantity is relatively insensitive to sea-quark masses. Consequently each of these quantities can be used to generate lattice spacings on an ensemble-by-ensemble basis. 

None of these quantities can be computed as accurately 
 as $r_1/a$ in simulations, but we can combine simulation results
for them with values for $r_1/a$ to obtain very accurate estimates
for the physical value of~$r_1$. Given $r_1$, the different values
of $r_1/a$ can be used to obtain accurate lattice spacings for each
of the simulations we discuss here and any other simulations where $r_1/a$ 
has been computed.

Of our three quantities, the $\eta_s$ decay constant gives the most accurate results. The $\eta_s$ is a fictitious meson, however, and so its ``experimental'' properties must be related to those of real mesons using simulations. The $\eta_s$ is particularly closely related to the $\pi$ and $K$ mesons. As we will show, its mass and decay constant can be accurately related to those of the $\pi$ and $K$ through a chiral analysis of simulation data for a variety of quark masses and lattice spacings. Such an analysis also gives an independent, fourth estimate of $r_1$.

We describe in section 2 the three primary methods we have 
used to obtain lattice spacings for a wide variety of simulations. 
Each can be used to generate an estimate for the physical value of $r_1$, given values 
of $r_1/a$.
In section 3 we combine the three analyses to generate a single, combined estimate for $r_1$. 
This can then be used to covert the $r_1/a$ values into a determination 
of $a$ on each ensemble. 
We also demonstrate how to determine the lattice spacing from the $\eta_s$ without 
using $r_1$. The two methods are compared and shown to agree in the $a \rightarrow 0$ limit. 
In Section~4 we give 
a value for $r_0$ derived from our value of $r_1$ for comparison to 
others using that parameter. In section 5 summarize our results.
Finally, we discuss the chiral analysis of decay constants and masses 
for the $\pi$, $K$ and their relation with those of the~$\eta_s$ meson 
in Appendices~A, B and~C. 

\section{Lattice calculation}

In Table~\ref{tab:params} we list the parameter sets for the 
different MILC ensembles of gluon configurations 
that we have used here, although not all ensembles were used in 
every lattice spacing determination. 

Values for the static-quark potential parameter $r_1/a$, in lattice units,
were determined by the MILC collaboration~\cite{milcreview}. 
They calculated the heavy quark potential 
by fitting Wilson loops of fixed spatial size as a function 
of lattice time. On the finest two sets of ensembles smeared time links were 
used to reduce statistical noise and a two-state exponential fit in time 
reduced the contamination from excited potentials. 
The heavy quark potential obtained was then fit as a function of spatial 
separation over the range between 0.2\,fm and 
0.7\,fm to a Cornell potential with the addition of 
corrections for lattice artifacts.
The point at which the condition for $r_1$ held was then determined from 
this fit. The errors given are statistical errors only, since discretisation 
effects are taken care of in our continuum extrapolations. 

In what follows we will combine these values for $r_1/a$ with estimates of the lattice spacing~$a$ determined using three different physical quantities to obtain estimates for the physical value of $r_1$ (that is, at zero lattice spacing and with correct sea-quark masses).

\subsection{$m_{\Upsilon^\prime}-m_\Upsilon$}
\label{sec:ups}

The calculation of the spectrum of mesons formed as bound states of 
bottom quarks and antiquarks has been an important 
test for lattice QCD. There are many radial and orbital 
excitations below threshold for strong decay and so many
gold-plated states, well-characterised experimentally. 
The radial and orbital excitation energies 
are almost identical for charmonium and bottomonium when 
spin-averaged~\cite{fnpdg08} 
and so rather insensitive to the heavy 
quark mass. Heavy-quark vacuum polarization effects are tiny 
and so can be safely neglected. This makes these systems very 
suitable for the determination of the lattice spacing~\cite{alan} 
and was one of the key calculations demonstrating the 
importance of including the effect of $u$, $d$ and $s$ sea quarks~\cite{ourlatqcd}. 
 
Here we improve on the calculations in~\cite{alan} which 
used results from MILC super-coarse, coarse and fine ensembles 
and compared ensembles with and without sea quarks. We 
study only ensembles including sea quarks but include also 
very coarse and superfine ensembles for a wider range of 
lattice spacing values. 

We calculate $b$-quark propagators on the MILC gluon 
field configurations using lattice NonRelativistic QCD (NRQCD)
which has been developed over many years to handle well the 
physics of heavy quark systems on the lattice~\cite{thackerlepage}. It makes a virtue 
of the nonrelativistic nature of bottomonium bound states
($v_b^2 \approx 0.1$ for the $\Upsilon$) by discarding the rest 
mass energy in favour of accurately handling typical 
momentum and energy scales inside the bound states. 
NRQCD can be matched to full QCD order by order in $v_b^2$ and 
$\alpha_s$. We work through $\cal{O}$$(v_b^4)$ in the nonrelativistic 
expansion and apply discretisation improvements through $\cal{O}$$(a^2)$ 
to $v_b^2$ terms and to chromomagnetic and chromoelectric field-dependent 
terms at $v_b^4$ (so that terms which induce fine structure are 
completely improved to $\cal{O}$$(a^4)$). An analysis of 
remaining systematic errors is given in~\cite{alan}.  

The NonRelativistic Hamiltonian that we use is given by~\cite{corrn}: 
 \begin{eqnarray}
 aH &=& aH_0 + a\delta H; \nonumber \\
 aH_0 &=& - \frac{\Delta^{(2)}}{2 aM_b}, \nonumber \\
a\delta H
&=& - c_1 \frac{(\Delta^{(2)})^2}{8( aM_b)^3}
            + c_2 \frac{ig}{8(aM_b)^2}\left(\bf{\nabla}\cdot\tilde{\bf{E}}\right. -
\left.\tilde{\bf{E}}\cdot\bf{\nabla}\right) \nonumber \\
& & - c_3 \frac{g}{8(aM_b)^2} \bf{\sigma}\cdot\left(\tilde{\bf{\nabla}}\times\tilde{\bf{E}}\right. -
\left.\tilde{\bf{E}}\times\tilde{\bf{\nabla}}\right) \nonumber \\
 & & - c_4 \frac{g}{2 aM_b}\,{\bf{\sigma}}\cdot\tilde{\bf{B}}  
  + c_5 \frac{a^2\Delta^{(4)}}{24 aM_b} \nonumber \\
 & & -  c_6 \frac{a(\Delta^{(2)})^2}{16n(aM_b)^2} .
\label{deltaH}
\end{eqnarray}
This is implemented in calculating $b$ quark propagators by evolving 
the $b$ quark Green's function on a single pass through the lattice 
using: 
\begin{eqnarray}
G(\vec{x},t+1) &=& (1-\frac{a\delta H}{2})(1-\frac{aH_0}{2n})^nU^{\dag}_{t}(x) \nonumber \\
        & & (1-\frac{aH_0}{2n})^n(1-\frac{a\delta H}{2}) G(\vec{x},t) 
\label{eq:evol}
\end{eqnarray}
with starting condition:
\begin{equation}
G(\vec{x},0) = \phi(x)\mathtt{1}.
\end{equation}
Here $\nabla$ is the symmetric lattice derivative and 
$\tilde{\nabla}$ is the improved derivative, $\tilde{\nabla}_k = \nabla_k - \nabla_k^{(3)}/6$.
$\Delta^{(2)}$ is the standard lattice discretisation of the 
second derivative $\sum_j \nabla_j^{(2)}$ and 
$\Delta^{(4)}$ is $\sum_j \nabla_j^{(4)}$.
$aM_b$ is the bare $b$ quark mass in lattice units. 

$\phi(x)$ is a real spatial smearing function which multiplies 
a unit matrix in color and spin space as the starting point for the quark 
propagator. 
The antiquark propagator for a given source is then the 
complex conjugate of the Green's function obtained from eq.~\ref{eq:evol}. 
When the quark and antiquark propagators 
are combined (with appropriate Pauli matrices for different 
$J^{PC}$~\cite{oldups}) into meson correlators, $\phi$ improves the overlap 
with particular ground and excited states for a better signal. 
This will be discussed further below. 

\begin{table}
\caption{\label{tab:params}
Ensembles (sets) of MILC configurations with gauge coupling $\beta$, 
size $L^3 \times T$ and sea 
mass parameters $m_{l}^\mathrm{asq}$ and $m_{s}^\mathrm{asq}$ used for this analysis. 
The sea 
ASQTAD quark masses ($l = u/d$) are given in the MILC convention where $u_0$ is the plaquette 
tadpole parameter. 
The lattice spacing values in units of $r_1$ after `smoothing'
are given in the third column~\cite{milcreview}. 
Sets 1 and 2 are `very coarse'; sets 3, 4 and 5, `coarse'; sets 6 and 7
`fine'; set 8 `superfine' and set 9 `ultrafine'. }
\begin{ruledtabular}
\begin{tabular}{llllllll}
Set & $\beta$ & $r_1/a$ & $au_0m_{l}^\mathrm{asq}$ & $au_0m_{s}^\mathrm{asq}$ & $L/a$ & $T/a$ \\
\hline
1 & 6.572 & 2.152(5) & 0.0097 & 0.0484 & 16 & 48 \\
2 & 6.586 & 2.138(4) & 0.0194 & 0.0484 & 16 & 48 \\
\hline
3 & 6.76 & 2.647(3) & 0.005 & 0.05 & 24 & 64  \\
4 & 6.76 & 2.618(3) & 0.01 & 0.05 & 20 & 64 \\
5 & 6.79 & 2.644(3) & 0.02 & 0.05 & 20 & 64 \\
\hline
6 & 7.09 & 3.699(3) & 0.0062 & 0.031 & 28 & 96  \\
7 & 7.11 & 3.712(4) & 0.0124 & 0.031 & 28 & 96  \\
\hline 
8 & 7.46 & 5.296(7) & 0.0036 & 0.018 & 48 & 144  \\
\hline
9 & 7.81 & 7.115(20) & 0.0028 & 0.014 & 64 & 192 \\
\end{tabular}
\end{ruledtabular}
\end{table}

$n$ is a stability parameter 
which is chosen to tame (unphysical) high momentum modes of 
the $b$ quark propagator which might otherwise cause the meson 
correlators to grow exponentially with time rather than 
fall. We used $n=4$ throughout this calculation instead of the 
value $n=2$ used earlier~\cite{alan}, so that we could work on finer 
lattices and keep the same value of $n$ for all ensembles. 
This means that discretisation errors are smoothly connected 
from one lattice spacing to another and the higher value of 
$n$ has the advantage of reducing some systematic errors.  

We use `tadpole-improvement'~\cite{lepmac} for all terms by
dividing the gluon fields $U_{\mu}$ by a factor $u_0$ when 
they are read in. $u_0$ is taken as $u_{0L}$, the value of the mean 
trace of the gluon field for that ensemble in lattice Landau gauge 
(where the trace is maximised). This removes, in a mean-field way, 
the disparity between lattice and continuum gluon fields induced 
by the fact that the lattice field is exponentially related 
to the continuum field. Single composite operators, such as 
$\Delta^{(4)}$,  are expanded out fully so that all cancellations 
between $U$ and $U^{\dag}$ are correctly tadpole-improved. 
This is not done for the $(1-{aH_0}/{2n})^n$ terms.  
Table~\ref{tab:upsparams} gives our parameters for the ensembles 
used in the $\Upsilon$ analysis. 

\begin{table}
\caption{	\label{tab:upsparams}
 Parameters used in our calculations of $b$ quark propagators 
and $b\overline{b}$ correlators on various MILC ensembles, numbered 
as in Table~\ref{tab:params}.  $M_ba$ is the bare $b$ quark mass in lattice units, 
$u_{0L}$ is the tadpole-improvement factor and the stability factor 
$n$ is taken as 4 everywhere. $n_{cfg}$ gives the number of gluon field
configurations used from the ensemble and $n_t$ is the number of 
time sources for $b$ quark propagators per configuration. $T$ is the 
time length in lattice units of the propagators. $a_0$ is the size parameter
for the quark smearing function $\phi_{es}(x)$ given in eq.~\ref{eq:smear}. 
}
\begin{ruledtabular}
\begin{tabular}{lllllll}
Set & $aM_b$ & $u_{0L}$ & $n_{cfg}$ & $n_t$ &  $T$ & $a_0$  \\
\hline
1 & 3.4 & 0.8218 & 631 & 24 & 32 & 0.83 \\
2 & 3.4 & 0.8225 & 631 & 24 & 32 & 0.83 \\
\hline
3 & 2.8 & 0.8362 & 2083 & 32 & 32 & 1.0 \\ 
4 & 2.8 & 0.8359 & 595 & 32 & 32 & 1.0 \\ 
\hline
6 & 1.95 & 0.8541 & 557 & 8 & 48 & 1.41 \\
\hline 
8 & 1.34 & 0.8696 & 698 & 8 & 48 & 2.0 \\
\end{tabular}
\end{ruledtabular}
\end{table}

In order to reduce statistical errors over our previous 
calculation we have investigated a number of improvements. 
The first was to look at different forms for the quark smearing 
$\phi(x)$. The simplest is a $\delta$ function but in addition 
we can take an arbitrary functional form for $\phi(x)$ provided that the 
gluon field configurations are gauge-fixed, at least on a
time-slice. The MILC configurations that we use here are fixed to Coulomb gauge. 
When a $b$ quark propagator from a $\delta$ 
source and a $\overline{b}$ propagator from a $\phi=f(x)$ source 
are combined a good overlap with a particular $\Upsilon$ 
state is expected when, in the language of a potential model,
 $f(x)$ is a good approximation to the wavefunction of that state.  
The ground state $\Upsilon(1S)$ will dominate all $1^{--}$ correlators 
eventually so that there is no advantage in including a smearing function
that gives a good overlap with that state~\cite{ps}. Instead, to obtain 
a good signal for the $2S-1S$ splitting, we concentrated on functions 
that had very small overlap with the 1S state, and therefore had 
better information about radial excitations.  A very 
good smearing for this was the function from~\cite{alan} called $\phi_{es}$:
\begin{equation}
\phi_{es}(r) = (2a_0-r) \exp(-r/(2a_0)).
\label{eq:smear}
\end{equation}
The size parameter, $a_0$, was tuned on coarse lattices to 
reduce the overlap of the correlator 
(known as the `ee' correlator, see below) with the ground state, as judged by 
the small amplitude of the correlator at large times when the ground
state dominates. $a_0$ was then scaled as appropriate to ensembles of 
different lattice spacing. Values are given in Table~\ref{tab:upsparams}. 
By combining $b$ and $\overline{b}$ propagators
from $\delta$ function sources and 
$\phi_{es}$ sources we are able to make up 3 different meson smearing 
functions: $l$ is from combining two $\delta$ sources; $e$ is from combining 
a $\phi_{es}$ source with a $\delta$ source and $E$ is from 
combining two $\phi_{es}$ sources (so that the composite meson smearing function is 
then the convolution of $\phi_{es}$ with itself). $l$, $e$ and $E$ smearing 
functions can also be applied at the sink to make a $3\times3$ matrix of 
correlators, with notation $ll$, $le$, $ee$ etc.  

We also used a random wall source for our $b$ quark propagators, taking a set 
of U(1) random numbers, $r$, with unit norm at every point on a time slice, 
one set for each color of the $b$ quark propagator. These 
were combined with the smearing functions $\phi$ so that 
\begin{equation}
G(\vec{x},0)_{c_1c_1} = \sum_{\vec{y}} \phi(|\vec{x}-\vec{y}|)r(c_1,\vec{y})\mathtt{1}_{spin}.
\end{equation}
When quark and antiquark propagators are combined together the 
random noise cancels except where the initial spatial sites are
the same and this effectively increases the number of meson 
correlators sampled. We find that the error on the ground 
state $\Upsilon$ energy is reduced by a factor of 3 on coarse 
lattices and 5 on fine lattices, when corrected to the same 
number of configurations. The excited state energy does not 
improve by the same factor, however. Indeed we found rather little 
improvement in the error on excited state energies which mirrors 
our experience with applying random wall sources to $B$ mesons~\cite{gregory}. 
The inference is that random wall sources are much less 
effective in situations where the degradation of the signal/noise 
is exponential. 
We calculate propagators from many different time sources 
(which we then average over) 
per configuration to improve statistical precision further. The details 
of numbers of configurations and time sources are collected
in Table~\ref{tab:upsparams}. 

As in~\cite{alan} we use a Bayesian fitting method~\cite{bayes} 
to fit the $3 \times 3$ matrix 
of hadron correlators to a multi-exponential form to extract 
the energies of states appearing in that correlator. 
This alows us to fit the entire correlator (i.e. for all time 
separations between source and sink), so making use of all 
the information contained in it. It also means that the fit 
results we obtain, for example for the ground state,
 include the effect of the higher excited states that are present 
in the correlator, and are not biassed by an attempt to fit only 
one or two states in a particular time window. 
The fitting function is 
\begin{equation}
G_{\mathrm{meson}}(n_{sc},n_{sk};t) = \sum_{k=1}^{n_{exp}}a(n_{sc},k)a^*(n_{sk},k)e^{-E_kt}.\label{eq:fit}
\end{equation}
where $a(n_{sc/sk},k)$ are the (real) amplitudes for state $k$ to appear in the 
smearings used at the source and sink of the correlator respectively. 

The Bayesian fitting method~\cite{bayes} allows a large number 
of exponentials to be used in the fit by constraining the way 
in which these exponentials can appear based on physical 
information. The simplest physical information is that the 
energies of states are ordered, and we implement this in the fit 
by taking the energy fit parameters as the natural 
logarithms of the ground state energy and of the energy splittings 
between adjacent states. 
On top of this we apply priors to the splittings between adjacent 
states that constrain them to be of order 500\,MeV with a 
width of a factor of two, i.e. between 250\,MeV and 1000\,MeV. 
Amplitudes are typically constrained around zero with a width 
of 1.0 (our composite meson smearing functions are normalised so that the spatial sum 
of their square is 1). We apply a cut on the range of eigenvalues 
present in the correlation matrix of $10^{-3}$ except for the high 
statistics calculation on the coarse 005/05 lattices where we use
$10^{-4}$. This reduces the number of degrees of freedom in the fit 
to between 120 and 170, with 208 in the coarse 005/05 fit. 
We obtain values for the $\Upsilon$ ground 
state energy and that of the first radial excitation, 
the $\Upsilon^{\prime}$ as a function of the number of exponentials in 
the fits. We demand 
a good $\chi^2$ and that the fit for 3 adjacent exponentials 
should agree both on the fitted values for 
the energies of interest $and$ on the errors.  
The ground state energy stabilises very quickly, but the 
first excited state is not generally stable until we reach 
8 exponentials. 
Fit results on the different MILC ensembles are then tabulated 
in Table~\ref{tab:upsresults} from 10 exponential fits. 

\begin{figure}
\begin{center}
\includegraphics[width=80mm]{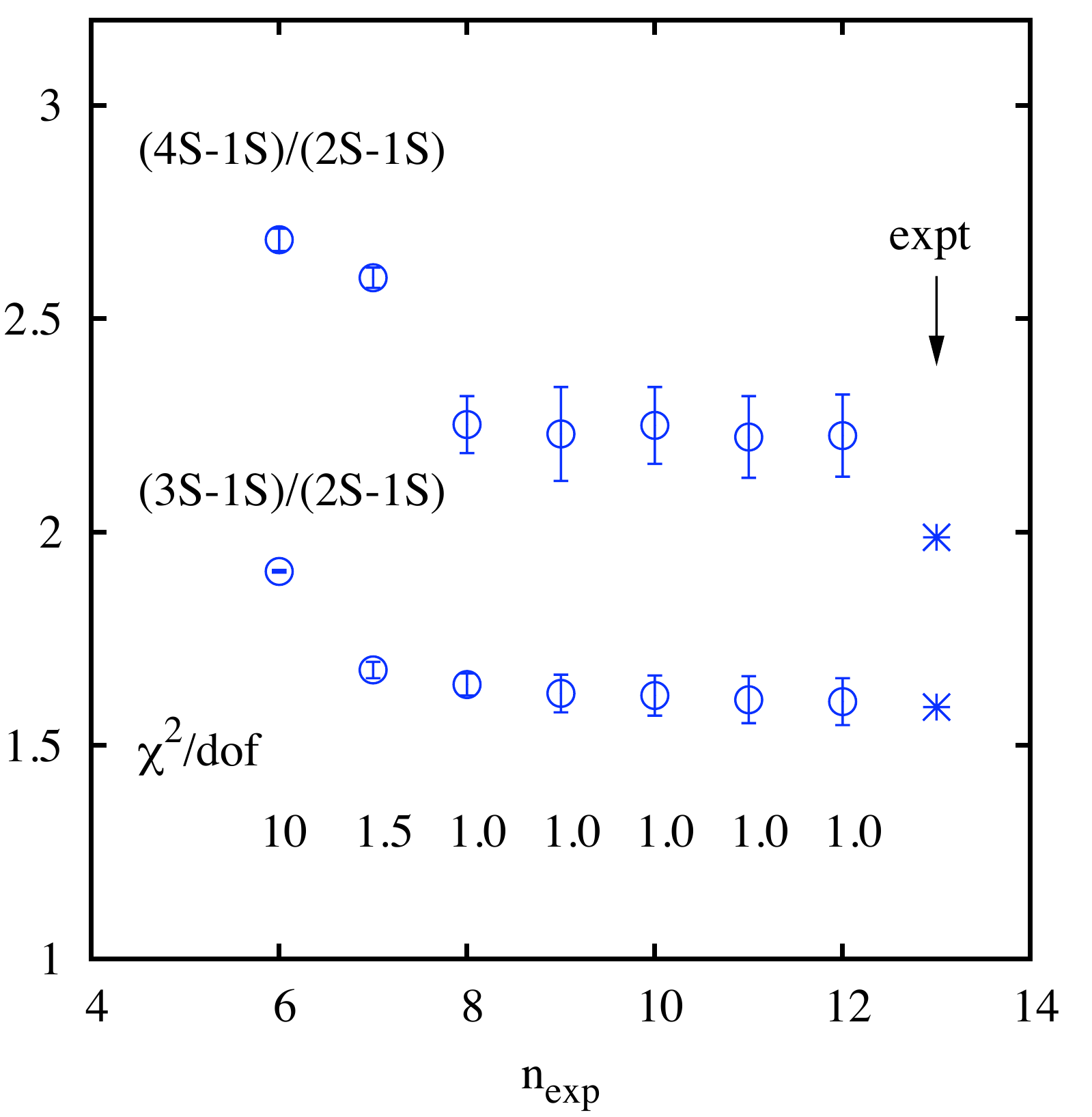}
\end{center}
\caption{Results for highly excited states from 
our fit to the $3\times 3$ matrix 
of $\Upsilon$ correlators from the coarse 005/05 (set 4) ensemble 
as a function of the number of exponentials included 
in the fit. The $\chi^2/dof$ is also shown - the fit had 208 
degrees of freedom. The results are stable from 9 to 12 
exponentials.}
\label{fig:upsfit}
\end{figure}

Figure~\ref{fig:upsfit} shows results from our highest 
statistics calculation on the coarse 005/05 lattices. 
Here we are able to obtain a good signal for even 
higher excited states than the $2S$. The plot shows 
the ratio of the $3S - 1S$ splitting and the $4S - 1S$ 
splitting to that of the $2S - 1S$. The $3S - 1S$ splitting 
is obtained to 3\% and in agreement with experiment. 
The $4S - 1S$ splitting is not very accurate even with 
the statistics available here. The result is slightly 
higher than experiment, but the $4S$ state is not gold-plated, 
decaying to $B\overline{B}$. This is not taken account 
of accurately in the lattice calculation and so we expect 
our result to be higher than experiment.  In our lower statistics calculations 
we do not generally have a significant signal for the $4S$  
and our $3S - 1S$ splitting has an error of between 5\% and 10\%. 

As discussed earlier, the excitation energies for 
bound states of heavy quarks are almost independent of the 
heavy quark mass, meaning that accurate tuning of this mass 
is not required for these splittings. Use of the random 
wall does, however, allow us to determine the meson 
energy as a function of meson momentum much more accurately 
than in previous calculations, and so the meson `kinetic' 
masses can be well determined. 
The meson mass in NRQCD must be determined from the meson 
dispersion relation because the zero of energy has an offset. 
The mass is then given by the difference in energy between 
mesons at zero momentum and momentum $pa$ on the lattice by~\cite{alan}:
\begin{equation}
Ma = \frac{p^2a^2 - (\Delta E a)^2}{2\Delta E a}.
\label{eq:kinmass}
\end{equation}
$\Delta E a$ is calculated by taking the difference in energy 
of the ground state 
from a simultaneous fit to $ll$ $\Upsilon$ correlators 
made from a standard (zero momentum) random wall as described 
above, and a random wall patterned with an appropriate Fourier 
factor to give a momentum of (1,0,0) 
to both quark and antiquark, so that the $\Upsilon$ has momentum 
(2,0,0) (and its permutations). Values obtained for the kinetic mass
on each of the MILC ensembles are given in Table~\ref{tab:upsresults}. 
They are tuned within 10\% of the experimental result of 9.46\,GeV~\cite{pdg08}. 

\begin{table}
\caption{	\label{tab:upsresults}
 Results for the ground state energy, $aE_1$, and 
radial excitation energy, $a(E_2-E_1)$ obtained from 
10 exponential fits of the form in equation~\ref{eq:fit} to a 
$3\times3$ matrix of $\Upsilon$ correlators as described 
in the text. The 4th column gives the $\Upsilon$ 
mass as determined from eq.~\ref{eq:kinmass}. Fewer 
configurations were used for this than for the full 
calculation (and given in Table~\ref{tab:upsparams}) 
in several cases. For set 3, 202 configurations 
were used and for set 8, 470 configurations.  
The 5th column gives the result for the lattice spacing 
from setting the $2S-1S$ splitting equal to the experimental 
value of 0.5630\,GeV~\cite{pdg08}. 
}
\begin{ruledtabular}
\begin{tabular}{lllllll}
Set & $aE_1$ & $a(E_2-E_1)$ & $aM$ & $a^\Upsilon$/fm \\
\hline
1 & 0.28775(8) & 0.4244(33) & 7.226(12) & 0.1488(12) \\
2 & 0.28814(8) & 0.4309(32) & 7.231(12) & 0.1510(11) \\
\hline
3 & 0.29330(3) & 0.3439(8) & 5.983(10) & 0.1205(3) \\ 
4 & 0.29261(6) & 0.3462(38) & 5.985(11) & 0.1213(13) \\ 
\hline
6 & 0.26618(5) & 0.2381(37) & 4.281(12) & 0.0835(13) \\
\hline 
8 & 0.24850(3) & 0.1679(14) & 3.050(18) & 0.0588(5) \\
\end{tabular}
\end{ruledtabular}
\end{table}

Table~\ref{tab:upsresults} gives results for the lattice
spacings $a^\Upsilon$ obtained by dividing the simulation results 
for $a(E_2-E_1)$ by the experimental value of~0.5630\,GeV for the splitting. Statistical errors 
are at the level of 1\%. 
Systematic errors arise from two sources, discretisation errors 
and missing higher-order relativistic correctiosn to the 
NRQCD action. The former can be removed by continuum extrapolation 
as long as they are well-behaved. The leading discretisation corrections
come from radiative corrections to existing terms in the action and 
can be calculated in perturbation theory. They 
have been shown to be small corrections in the region of $aM_b$ in 
which we work, and relatively independent of $aM_b$~\cite{morning2}. 
Relativistic corrections 
survive the continuum limit and are the main source of systematic 
error for this method.  
They were estimated in~\cite{alan} at 0.7\% on the coarse and 
0.6\% on the fine lattices, so we include an overall systematic 
error of 0.7\% in our error analysis here. 

One ingredient missing from our calculation and present in the 
experimental world is electromagnetism. This is then another 
possible source of systematic error. From a potential model 
calculation we estimate the shift in the $2S-1S$ splitting to be 
less than 1\,MeV from the Coulomb interaction between $b$ and $\overline{b}$ 
(the electromagnetic self-interaction is included in the $b$ quark mass). 
At less than 0.2\%, this is negligible. 

To extract a physical value for the static-quark potential parameter $r_1$, we must combine the lattice spacings $a_i^\Upsilon$ in Table~\ref{tab:upsresults} with the corresponding values of $(r_1/a)_i$ in Table~\ref{tab:params}, and extrapolate to zero lattice spacing, correcting the sea-quark masses. We do this by fitting $(r_1/a)_i a_i^\Upsilon$ from the $i$th ensemble to a formula for the effective $r_1$ corresponding to $m_{\Upsilon^\prime}-m_{\Upsilon}$:
\begin{align}
	\label{eq:r1-ups}
	r_1^{\Upsilon}(a,
			\delta m_l^\mathrm{sea},&\delta m_s^\mathrm{sea}) =
			 r_1
			\\ \nonumber
			&\times \left(1+c_\mathrm{sea}^\Upsilon \frac{2 \delta 	
			m_l^\mathrm{sea}+\delta m_s^\mathrm{sea}}{m_s} \right) 
			\\ \nonumber
			& \times \left(1 
			+ \sum_{j=1}^4 c_j^\Upsilon (a/r_1)^{2j}
			\right),
\end{align}
where $r_1$ (the extrapolated value), $c_\mathrm{sea}^\Upsilon$ and $c_j^\Upsilon$ are the parameters tuned by the fit. Here the $\delta m^\mathrm{sea}$ are differences between the sea-quark masses used in the simulation and the correct masses for $l=u/d$ and $s$~quarks (see Appendix~\ref{sec:sea-quark-masses}).

We have included twice as many terms as we need in the expansion in $a/r_1$; taking half as many terms gives essentially identical results. We are able to retain higher-order terms because we include Bayesian priors in our fit for each expansion coefficient used\,---\,that is, we include an initial estimate for each parameter. Each prior functions as an additional piece of input data in the fit, thereby guaranteeing that we always have more fit data than parameters, no matter how many parameters we choose to keep. In Bayesian fits like ours it is important to keep more parameters rather than fewer, not because they improve the fit but rather because they help us avoid underestimating our extrapolation errors. Here we used priors $c_j^\Upsilon = 0(1)$, $c_\mathrm{sea}^\Upsilon=0.0(1)$, both of which are broader (\emph{i.e.,} more conservative) than suggested by the empirical Bayes criterion~\cite{bayes}. Setting $c_\mathrm{sea}^\Upsilon$ to zero has negligible impact, so this parameter is not really necessary. We also take a very broad prior for $r_1$ that encompasses all current estimates: $r_1=0.315(10)$\,fm. It has little impact on our final errors.

\begin{table}
	\begin{ruledtabular}
		\caption{\label{tab:r1-error}
		Major sources of uncertainty in physical $r_1$ values 
		obtained from simulation results for $m_{\Upsilon^\prime}-m_\Upsilon$, 
		$m_{D_s}-m_{\eta_c}/2$, $f_{\eta_s}$, and from an analysis that fits 
		all three types of data simultaneously.}
	\begin{tabular}{rrllll}
			&& $\Upsilon$ & $D_s$ & $f_{\eta_s}$ & combined \\
			\hline
	           $a^2$ extrapolation &&  0.4\%  &  0.8\%  &  0.1\%  &  0.2\% 	\\
	          $m_s$ extrapolations &&  --	  &  0.0    &  0.0	  &  0.0   \\
	           $r_1/a$ uncertainty &&  0.2	  &  0.2    &  0.1	  &  0.1   \\
	  initial uncertainty in $r_1$ &&  0.6	  &  0.8    &  --	  &  --   \\
	   $\pi$-$K$-$\eta_s$ analysis &&  --	  &  --     &  0.8	  &  0.6   \\
	            statistical errors &&  1.1	  &  0.4    &  0.2	  &  0.3   \\
	         sea-quark mass tuning &&  0.1	  &  0.2    &  0.0	  &  0.1   \\
	      overall systematic error &&  0.6	  &  1.1    &  --	  &  0.2   \\
	\hline
	                        Total  && 1.4\%	  & 1.6\%   & 0.9\% & 0.7\% \\
	\end{tabular}
\end{ruledtabular}
\end{table}

Fitting our data, we obtain a final value for the physical $r_1$ from the upsilon simulations of:
\begin{equation}
	\label{r1-ups}
	r_1 = 0.3091(44)\,\mathrm{fm}\quad\mbox{(from $m_{\Upsilon^\prime}-m_\Upsilon$).}
\end{equation}
The fit is excellent, with a $\chi^2$ per degree of freedom of~0.2. We show plots in Section~III. The main sources of error in this result are listed in the $\Upsilon$-column of Table~\ref{tab:r1-error}; most of the error is due to statistical errors from the Monte Carlo simulation. Improvement will require much higher statistics 
on the fine and superfine lattices. 

\subsection{$m_{D_s} - m_{\eta_c}/2$}
\label{sec:ds}

A useful mass difference in the charm sector is that between 
the $D_s$ meson and one half of the mass of a low-lying 
$c\overline{c}$ state. We choose the $\eta_c$ rather than the $J/\psi$ because it is 
the easiest for us to calculate. This splitting 
has the experimental value 0.4784(7)\,GeV~\cite{pdg08} which 
changes to 0.672(2)\,GeV when the charm quark is replaced by a bottom quark. 
So the sensitivity to the heavy-quark mass, while 
stronger than for the heavy-onium splittings, is still 
rather mild. Both the $D_s$ and the $\eta_c$ are ground state mesons 
and do not have the poor signal/noise issues that the $\Upsilon(2S)$ 
state had in the previous subsection. Two quark masses are involved 
in the $D_s - \eta_c/2$ splitting, however, and this makes the 
tuning rather complicated. As a result we have only done this on 
two ensembles - coarse set 4 (see Table~\ref{tab:params}) using 
595 configurations with 2 time sources for propagators on each configuration  
and fine set 6 using 566 configurations with 4 time sources each.

Following the development of the HISQ action~\cite{hisq}
it is now possible to handle charm quarks accurately 
with a relativistic action in lattice QCD. The success of 
this action is demonstrated by an accurate (7\,MeV) determination, 
that agrees with experiment, of the 
masses of $D$ and $D_s$ meson when the charm quark mass is 
fixed from the $\eta_c$~\cite{fds}. This has not been possible with any 
other discretisation of QCD for charm quarks. Here we are 
essentially inverting this calculation to use $m_{D_s} - m_{\eta_c}/2$ 
to determine a value for the lattice spacing, simultaneously 
requiring that the mass of the $\eta_c$ and the $\eta_s$ 
(the fictitious pseudoscalar particle made of an $s$ quark-antiquark 
pair, see subsection C) be correct. We use the HISQ action as described in~\cite{hisq}
except that we simplify the tuning of the coefficient of the Naik 
term (that corrects for $a^2$ errors) so that it is 
correct as a function of $ma$ at tree level. In~\cite{hisq} it 
was shown that a nonperturbative tuning of this coefficient 
gave results very similar to the tree-level result and so it is much simpler to
take the tree-level result at each value of the quark mass.  
The difference between tree-level and nonperturbative tuning 
of the coefficient is a small discretisation error at 
relatively high order, and it will be taken care of in our 
continuum extrapolation. 

\begin{figure}
\begin{center}
\includegraphics[width=80mm]{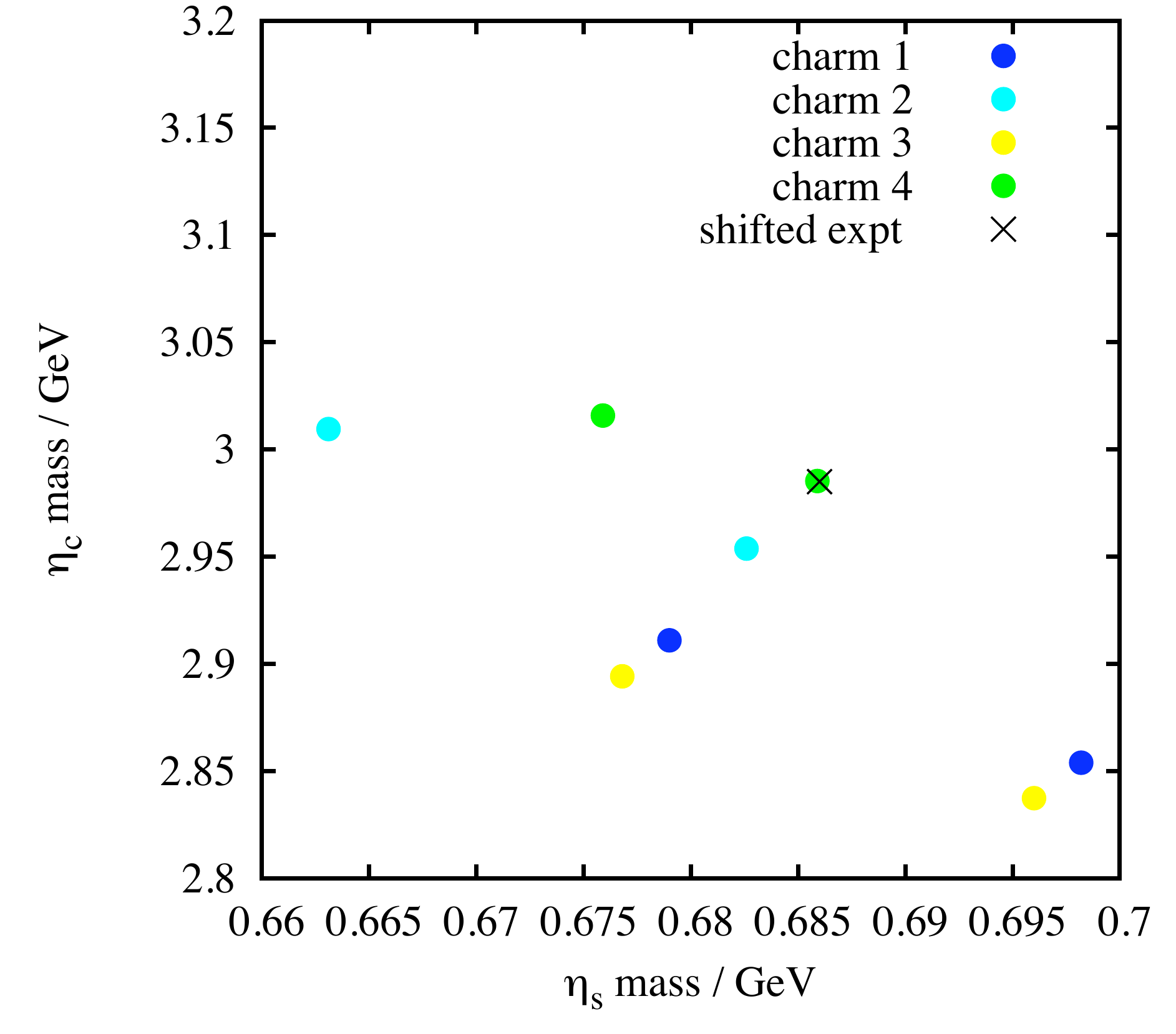}
\end{center}
\caption{Results for $m_{\eta_c}$ vs $m_{\eta_s}$ for different 
charm and strange quark masses on the coarse 01/05 ensemble (set 4). 
Points corresponding to different charm quark masses are given 
in different colors. Several different strange quark masses are given 
for each charm quark mass.  
The lattice spacing is determined from $m_{D_s}-m_{\eta_c}/2$. Note that 
the experimental point is shifted to allow for electromagnetic effects missing 
from our calculation, as described in the text. }
\label{fig:ds-etac}
\end{figure}

We proceed by calculating $\eta_c$, $\eta_s$ and $D_s$ correlators 
for several different combinations of bare quark masses for charm 
and strange quarks. We use random wall sources as for the $b$ quarks 
in the previous subsection, which improves the statistical error on 
the ground state masses that we need here significantly. No smearing function 
is necessary. Each of the 
correlators is fit to an appropriate multi-exponential form including 
oscillating states for the $D_s$~\cite{fds}. 
\begin{equation}
G_{\mathrm{meson}}(t) = \sum_{k=1}^{n_{exp}}a_k(e^{-M_kt}+e^{-M_k(T-t)}),\label{eq:fitstagg}
\end{equation}
for $\eta_c$ and $\eta_s$ and 
\begin{eqnarray}
G_{\mathrm{meson}}(t) &=& \sum_{k=1}^{n_{exp}}a_k(e^{-M_kt}+e^{-M_k(T-t)}) \\
&+& \sum_{ko=1}^{n_{exp}}a_{ko} (-1)^ta_{ko}(e^{-M_{ko}t}+e^{-M_{ko}(T-t)})\nonumber \label{eq:fitstaggosc}
\end{eqnarray}
for $D_s$. As before, standard Bayesian fitting techniques are used~\cite{bayes} 
and results are taken from fits with $n_{exp}=4$, where we find the results, 
and their errors, are stable to changes in $n_{exp}$. 

For each combination it is then possible to 
plot the value of $m_{\eta_c}$ against that of $m_{\eta_s}$ using the 
lattice spacing from $m_{D_s}-m_{\eta_c}/2$ and interpolate
to experiment.  A plot showing our results on the coarse 01/05 (set 4) 
ensemble is given in Figure~\ref{fig:ds-etac}.  Interpolation to the 
matching point is relatively simple because the $D_s$ mass is linear 
in $m_sa$ and $m_ca$, the $\eta_c$ mass is linear in $m_ca$ and 
the square of the $\eta_s$ mass is linear in $m_sa$ for small changes 
in the quark masses. As can be 
seen from the Figure it is possible to pinpoint the matching point precisely 
and then confirm it with an additional calculation.  Statistical errors 
below 0.5\% are achievable. 

The experimental values to be used here must take into account that the 
lattice calculation is missing the electromagnetism of the real world.  
This gives a significant shift to $m_{D_s}-m_{\eta_c}/2$ because 
the $D_s$ is electrically charged and the $\eta_c$ is neutral, so their 
masses shift in opposite directions and the shifts add together. We estimate 
the electromagnetic shift to the $D_s$ mass to be 1.4\,MeV, and the shift
to the $\eta_c$ mass to be -2.6\,MeV. In addition the $\eta_c$ in the 
real world can annihilate to gluons but not in our calculation and 
we estimate the shift from this effect to be -2.4\,MeV~\cite{hisq}. 
This shifts $m_{D_s} - m_{\eta_c}/2$ from the real world to 0.4745\,MeV 
for comparison to our calculation. We take the systematic error on 
this value to be one half of the shift we apply: that is, 2\,MeV. This 
0.5\% error gives a 1.5\% systematic error on the values of the lattice 
spacing that we obtain, unfortunately dominating our statistical errors.  
The experimental mass for the $\eta_c$ becomes 2.985\,GeV after applying 
the shifts above and the $\eta_s$ mass is taken as 0.686\,GeV 
(see Appendix~\ref{sec:chiral1} and Eq.~(\ref{eq:chiralresults})). 

Table~\ref{tab:dsetac} gives results on the coarse (set 4) and fine (set 6) lattices
from this method.  
The $\eta_s$ correlators are a subset of those used in subsection C. 
The results for $m_{\eta_s}$ are very slightly different in the 
two cases because of the different fitting strategy employed. 

\begin{table}
	\caption{\label{tab:dsetac} Results for $\Delta = m_{D_s}-m_{\eta_c}/2$ in lattice 
units from different charm and strange HISQ quark masses on ensembles 4 and 6. 
The corresponding lattice spacing, 
at the tuned point where $m_{\eta_c}$ and $m_{\eta_s}$ agree with their 
physical values, are given in the bottom row of each section of the table. 
Errors shown are from statistics and extrapolation only.  }
	\begin{ruledtabular}
\begin{tabular}{cccccc}
set 4 & & & & & \\
\hline
$m_ca$ & $m_{\eta_c}a$ & $m_sa$ & $m_{\eta_s}a$ &  $a\Delta$ & $a$/fm \\
0.72 & 1.98114(15) & 0.06 & 0.45787(23) & 0.3180(5)   & \\
0.753 & 2.04293(10) & 0.06 & 0.45787(23) & 0.3214(5) & \\
0.753 & 2.04293(10) & 0.063 & 0.46937(24) & 0.3247(5) & \\
\hline
& tuned & & & 0.3247(5) & 0.1350(2) \\ 
\hline
\hline
set 6 & & & & & \\
\hline
$m_ca$ & $m_{\eta_c}a$ & $m_sa$ & $m_{\eta_s}a$ &  $a\Delta$ & $a$/fm \\
0.44 & 1.33816(7) & 0.0358 & 0.30332(12) & 0.21244(23) & \\
0.44  & 1.33816(7)  &  0.0382 & 0.31362(14) & 0.21535(22) & \\
0.45 & 1.35934(7) & 0.0358 & 0.30332(12) & 0.21350(24) & \\
0.45 & 1.35934(7) & 0.0382 & 0.31362(14) & 0.21640(23) & \\
\hline 
& tuned & & & 0.2174(5) & 0.0904(2) \\ 
\end{tabular}
\end{ruledtabular}
\end{table}

Given the lattice spacings in Table~\ref{tab:dsetac}, we can again combine them with the corresponding values for $r_1/a$ from Table~\ref{tab:params} to obtain effective values $r_1^{D_s}$ that we can extrapolate to zero lattice spacing. We do this using the same parameterization and priors for $r_1^{D_s}$ as we did for $r_1^\Upsilon$ in the previous section, except that here we allow for less dependence on the sea-quark masses since that is what our previous simulations have shown~\cite{fds}. We take $c_\mathrm{sea}=0.00(1)$ as a prior. With only two data points, the fit is almost trivial, giving
\begin{equation}
	\label{r1-Ds}
	r_1 = 0.3157(53)\,\mathrm{fm}\quad\mbox{(from $m_{D_s}-m_{\eta_c}/2$),}
\end{equation}
which agrees well with our estimate from the $\Upsilon$ but is less accurate.
The main sources of error in this result are listed in the $D_s$-column of Table~\ref{tab:r1-error}; the largest source of error is the overall systematic error~\cite{fnsyst}. 
The systematic error could be improved slightly by using 
the $J/\psi$ instead of the $\eta_c$ to avoid the sizeable mass shift from annihilation to gluons and its 
uncertainty. In addition a more accurate understanding of electromagnetic mass shifts, with quantitative 
tests on the lattice would help (see, for example,~\cite{milcem}).

\subsection{$f_{\eta_s}$}
\label{sec:fetas}

The $\eta_s$ is a fictitious pseudoscalar meson.
It is like the pion and kaon, but with valence $s\bar{s}$~quarks. In the real world the valence $s\bar{s}$~state mixes with~$u\bar{u}$ and~$d\bar{d}$, through valence quark-antiquark annihilation, to form the $\eta$ and $\eta^\prime$ mesons. By omitting valence quark-antiquark annihilation from our simulation, we obtain the $\eta_s$ instead. This meson is easily studied, in lattice QCD, using simulations and, in the continuum, using partially-quenched chiral perturbation theory~\cite{sharpe-pqX}. In Appendix~\ref{sec:chiral1} we show how to determine its mass and decay constant from simulation and experimental data for pions and kaons, using chiral perturbation theory.
We are able to determine both parameters to within about~0.5\%.

Given an accurate physical value, the $\eta_s$ mass is the easiest quantity to use for tuning the $s$-quark mass in lattice simulations. It is significantly simpler to use than the $K$~mass since $m_{\eta_s}$, unlike $m_K$, is only weakly dependent upon the $u/d$ mass and therefore requires only minimal chiral extrapolation. This is because $u/d$ quarks enter only in the sea for this meson. Another advantage of the $\eta_s$ is that it is much less expensive to simulate than the~$K$.

Given a tuned $s$-quark mass, $f_{\eta_s}$ is much more useful for tuning the lattice spacing than either $f_K$ or~$f_\pi$. Again, this is because it is almost independent of the $u/d$ mass (and because it is much less expensive to compute). We have computed both the decay constant and the mass for the $\eta_s$ for a variety of $s$-quark masses for all of our lattice parameter sets. The results are given in Table~\ref{tab:etas}.

\begin{table}
	\caption{\label{tab:etas}
	Simulation results for the $\eta_s$ mass $m_{\eta_s}$ 
	and decay constant $f_{\eta_s}$
	for several lattice parameter sets (see Table~\ref{tab:params}) 
	and $s$-quark masses $am_s$. We also list the number of gauge field 
 configurations and time sources per configuration used. }
	\begin{ruledtabular}
	\begin{tabular}{clccc}
		Set & $am_s$ & 
			$af_{\eta_s}$ & $am_{\eta_s}$ & $n_{cfg}\times n_t$ \\ \hline
		1 & 0.066 &  0.1429(4) & 0.5250(6) & $631 \times 2$\\
		 & 0.08 &  0.1485(4) & 0.5782(6) & $631 \times 2$ \\
		2 & 0.066 & 0.1436(4) & 0.5248(6) & $631 \times 2$ \\ \hline
		3 & 0.0537 & 0.1144(2) & 0.4310(4) & $518 \times 2$ \\
		4 & 0.0546 & 0.1160(3) & 0.4367(5) & $595 \times 2$ \\
		 & 0.05465 &  0.1160(3) & 0.4369(5) & $595 \times 2$ \\
		 & 0.06 &  0.1182(4) & 0.4580(5) & $595 \times 2$ \\
		5 & 0.0525 & 0.1149(4) & 0.4259(6) & $460 \times 2$ \\
		 & 0.0556 & 0.1161(4) & 0.4384(6) & $460 \times 2$ \\ \hline
		6 & 0.0358 & 0.0806(2) & 0.3035(3) & $566 \times 4$ \\
		 & 0.0366 & 0.0810(2) & 0.3069(3) & $566 \times 4$ \\
		 & 0.0382 & 0.0817(2) & 0.3137(3) & $566 \times 4$ \\
		7 & 0.03635 & 0.0811(2) & 0.3050(4) & $265 \times 4$ \\ \hline
		8 & 0.024 & 0.0556(1) & 0.2120(2) & $218 \times 4$ \\ \hline
		9 & 0.0165 & 0.0408(1) & 0.1548(1) & $200 \times 2$ \\
		 & 0.018 & 0.0417(2) & 0.1621(2) & $101 \times 1$ \\
	\end{tabular}
	\end{ruledtabular}
\end{table}

As discussed earlier, we used the HISQ formalism for the valence $s$~quarks in our analysis, together with the MILC gluon configurations described in Table~\ref{tab:params}. We analyzed the $\eta_s$ created by the partially conserved axial-vector current in the HISQ formalism, so that the decay constant is automatically correctly normalized, with no need for further renormalization constants. We used random-wall sources when computing quark propagators, as described earlier for $b$-quark propagators, and used sources on several time slices for each configuration, to increase statistics, see Table~\ref{tab:etas}. 

We extracted masses and decay constants from the meson correlators by fitting the middle 40\% of the $t$~range 
to a single exponential. This is less sophisticated than our approach to fitting correlators in previous subsections, 
but it simplifies the analysis of statistical correlations between different results coming from the same ensemble (with different $s$-quark masses). We get identical results if we use instead results from multi-exponential fits, ignoring correlations. The $\chi^2$~per degree of freedom of our fits was larger than one for some ensembles, possibly because of lower statistics. To be conservative, we doubled the statistical errors everywhere (giving the results in Table~\ref{tab:etas}), 
resulting in excellent~$\chi^2$s.

In analyzing our simulation results for $(af_{\eta_s})_i$ and $(am_{\eta_s})_i$, we need to account for three systematic effects. First none of the simulations has precisely the correct $s$-quark mass~$m_s$. We did simulations at multiple values of $m_s$ so that we could interpolate. The lattice spacing cancels out in the ratio $(af_{\eta_s})_i/(am_{\eta_s})_i$; we in effect vary~$m_s$ until this ratio has the correct continuum value, obtained from our chiral analysis (see Appendix~\ref{sec:chiral1}). 

The second important systematic effect is that our simulations have finite-lattice-spacing errors. We model dependence on the lattice spacing using a power series in $(a/r_1)^2$. A final, but much less important systematic is that the sea-quark masses are not quite right in our simulations. We did simulations using several different sea-quark masses so that we could correct for this dependence, which, as discussed above, we expect (and find) to be very small. Other systematic errors are negligible. In particular, finite-volume corrections for the $\eta_s$ are no larger than~0.1\% in our simulations.

We account for these systematic effects by fitting our results $(af_{\eta_s})_i$ from the simulation using ensemble set~$i$, with each $s$~mass, to:
\begin{equation}
	(a/r_1)_i\, r_1^{\eta_s} f_{\eta_s}^\mathrm{lat}(a_i,x_{\eta_s}),
\end{equation}
where again values for $(r_1/a)_i$ come from Table~\ref{tab:params}.
This formula defines $r_1^{\eta_s}$, which is the effective value of $r_1$ implied (for each ensemble set) by our data for the $\eta_s$~decay constant and mass. We parameterize $r_1^{\eta_s}$ the same way we parameterized $r_1^\Upsilon$ and $r_1^{D_s}$:
\begin{align}
	\label{eq:r1-etas}
	r_1^{\eta_s}(a,
			\delta m_l^\mathrm{sea},&\delta m_s^\mathrm{sea}) =
			 r_1
			\\ \nonumber
			&\times \left(1+c_\mathrm{sea}^{\eta_s} \frac{2 \delta 	
			m_l^\mathrm{sea}+\delta m_s^\mathrm{sea}}{m_s} \right) 
			\\ \nonumber
			& \times \left(1 
			+ \sum_{j=1}^4 c_j^{\eta_s} (a/r_1)^{2j}
			\right)
\end{align}
where again $r_1$ is the physical value.
Function $f_{\eta_s}^\mathrm{lat}(a,x_{\eta_s})$ models the $s$-quark mass dependence of the decay constant where
\begin{equation}
	\label{eq:fmetas}
	x_{\eta_s} \equiv \left(\frac{(am_{\eta_s})_i}{(af_{\eta_s})_i} \, 	
	\frac{f_{\eta_s}}{m_{\eta_s}}\right)^2 - 1
\end{equation}
is a measure of difference between the correct $s$~mass and the $s$~mass used in the simulation to produce~$(af_{\eta_s})_i$ and~$(am_{\eta_s})_i$. We parameterize $f_{\eta_s}^\mathrm{lat}$ as follows:
\begin{equation}
	\label{eq:fetas}
	f_{\eta_s}^\mathrm{lat}(a,x_{\eta_s}) = f_{\eta_s} + \sum_{k=1}^{4} d_k 
	x_{\eta_s}^k. 
\end{equation}
We allow the first two terms in the expansion to depend upon the lattice spacing by taking
\begin{equation}
	d_k \equiv d_{k0} + d_{k1} (a/r_1)^2
\end{equation}
for $j=1,2$; lattice-spacing dependence in the higher-order terms would have negligible effect (as do the higher-order terms themselves, as it turns out).

Again we have included twice as many terms as we need in the expansions in $a/r_1$ and~$x_{\eta_s}$; taking half as many terms in both cases gives essentially identical results. Here we used priors $c_j^{\eta_s} = 0(1)$ and $c_\mathrm{sea}^{\eta_s}=0.0(1)$, as before, and $d_{kl}=0.0(5)$. Again all priors are somewhat broader (that is, more conservative) than suggested by the empirical Bayes criterion~\cite{bayes}. 

The other parameter varied in the fit is the continuum/physical $r_1$ in Eq.~(\ref{eq:r1-etas}). We tried two different priors for this parameter. First we took the very broad prior, $0.315(10)$\,fm, we used for the other quantities. We also fit using the $r_1$~result from our chiral analysis of $f_\pi$ and $f_K$ in Appendix~\ref{sec:chiral1} (Eq.~(\ref{eq:chiralresults})). These two choices give results that differ by only a tenth of a standard deviation, which is negligible. We use the latter choice for our results below. We also take the values for $f_{\eta_s}$ and $(f_{\eta_s}/m_{\eta_s})$ used in Eqs.~(\ref{eq:fetas}) and~(\ref{eq:fmetas}) from our chiral analysis as described in Appendix~\ref{sec:chiral1}. 

Our final result for the continuum value for $r_1$ in this analysis is:
\begin{equation}
	\label{r1-etas}
	r_1 = 0.3148(28)(5)\,\mathrm{fm}\quad\mbox{(from $f_{\eta_s}$),}
\end{equation}
where, as discussed in Appendix~\ref{sec:chiral1}, the second error corresponds to uncertainty about finite-volume corrections in the chiral analysis. The fit is excellent, with a~$\chi^2$ per degree of freedom of~0.4. The main sources of error in this result are listed in the $f_{\eta_s}$-column of Table~\ref{tab:r1-error}; the largest source of error is uncertainty in the physical values of $f_{\eta_s}$ and $m_{\eta_s}$ from the $\pi$-$K$-$\eta_s$ chiral analysis.

\section{Two Recipes}
\label{sec:combined}
\begin{figure}
	\begin{center}
		\includegraphics{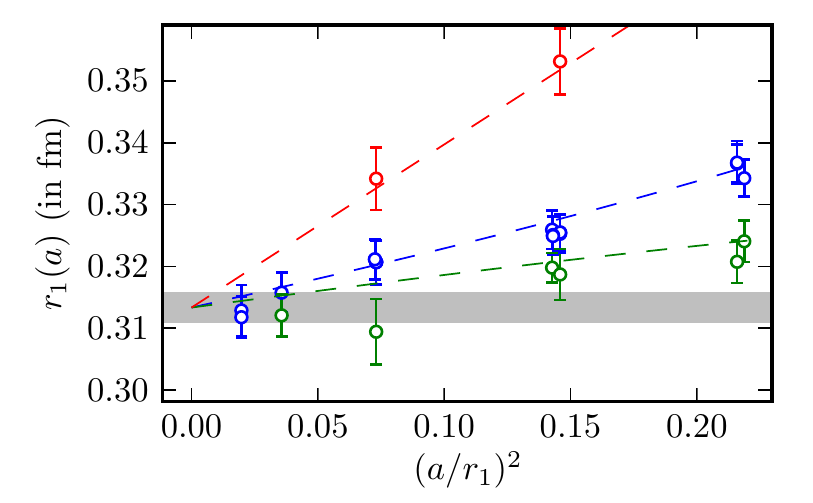}
		\vspace{-3ex}
	\end{center}
	\caption{Simulation results for the effective $r_1$ 	
	obtained from $m_{D_s}-m_{\eta_c}/2$ (top), $f_{\eta_s}$ (middle), and 	
	$m_{\Upsilon^\prime}-m_\Upsilon$ (bottom) 
	are plotted versus $(a/r_1)^2$ for 
	various values of the sea-quark mass. The lines show the tuned
	fit functions from our simultaneous 
	fit to all three sets of simulation results. We used the fit functions
	to correct the simulation data points for the sea-quark masses; data 
	points and lines are for $\delta m_q^\mathrm{sea}=0$.
	The gray band is the
	continuum value obtained from the fit: 	
	$r_1=0.3133(23)$\,fm.}
	\label{fig:r1-all}
\end{figure}

Two accurate recipes for setting the lattice spacing follow from the analysis in the previous section. The first requires that the static-quark potential be computed in the simulation, and a value for $r_1/a$ extracted from the results. This has been done accurately by the 
MILC collaboration for their ensembles and we use their numbers. 
$r_1/a$ can then be converted to a value for the lattice spacing by dividing into the physical value of~$r_1$. 
In the previous section, we did separate determinations of $r_1$ using simulation results for the upsilon and $D_s$ mass splittings, and for the $\eta_s$~decay constant. For each we extracted effective values of $r_1$ for each lattice ensemble and parameter set; and we extrapolated to the continuum to obtain physical values for~$r_1$. We have also done a joint analysis of all three sets of simulation results which is identical to what we did for each separately, but requiring that each fit use the same physical~$r_1$\,---\,that is, we require all three to agree on the final value for $r_1$. This analysis also implicitly includes the $r_1$ result from our chiral analysis of $f_\pi$ and $f_K$ since we use that value as the input prior for the combined analysis. When we do this we obtain the following final result, where again the second error is due to uncertainties in finite-volume corrections to the chiral analysis (see Appendix~\ref{sec:chiral1}):
\begin{equation}
	r_1 = 0.3133(23)(3)\,\mathrm{fm} \quad \mbox{(combined)}
\end{equation}
The fit is excellent with a $\chi^2$ per degree of freedom of~0.4. Figure~\ref{fig:r1-all} shows $r_1$~values from all three simulations plotted against the square of the lattice spacing. The sources of error in this combined analysis are summarized in the last column of Table~\ref{tab:r1-error}. The $a^2$ dependence in that figure is all relative to $a^2$ dependence in the $r_1/a$ values obtained from the static-quark potential. Thus the upsilon analysis has $a^2$ errors most similar to those in the static-quark potential's $r_1/a$, while the $D_s$ analysis has errors least like those coming from the static-quark potential. There is no way to tell which of these quantities has the smallest absolute finite-$a$ errors from just this simulation data; all that we can say is that they are consistent with each other in the continuum limit.

The second recipe for determining the lattice spacing for a particular configuration set 
requires only the evaluation of the mass and decay constant for the 
$\eta_s$ (see Section~\ref{sec:fetas}) on those configurations; 
there is no need for the static-quark potential in this recipe. 
Lattice results for $af_{\eta_s}$ are fit to the formula
\begin{equation}
	\label{eq:fetas2}
	af_{\eta_s}^\mathrm{lat} = a f_{\eta_s}\left(
	1 + c_1 x_{\eta_s} + c_2 x_{\eta_s}^2 \right),
\end{equation}
where, as before,
\begin{equation}
	x_{\eta_s} = \left(\frac{f_{\eta_s}}{m_{\eta_s}}\,
	\frac{am_{\eta_s}^\mathrm{lat}}{af_{\eta_s}^\mathrm{lat}}\right)^2 - 1,
\end{equation}
and $a$, $c_1$, and $c_2$ are fit parameters. Physical values for the mass and decay constant, $m_{\eta_s}$ and $f_{\eta_s}$, are again taken from Eqs.~(\ref{eq:chiralresults}) in Appendix~\ref{sec:chiral1}. Our simulations indicate that  $c_1=0.33(5)$ and $c_2=0.0(5)$ are good priors for these parameters; further terms in the $x_{\eta_s}$~expansion are unnecessary provided $x_{\eta_s}$~is small (it is less than 0.06 for our data). The  lattice spacing for the particular configuration set under study is then an output from the fit. 

In a typical simulation one guesses a value for the bare $s$-quark mass, $am_s$ in lattice units, to use in the quark action. Provided this is close enough to the correct value,  a fit of the $\eta_s$ results from this single mass is enough to generate an accurate lattice spacing. Doing simulations with two or more $s$-quark masses improves the result. 

The correct value for $am_s$ can also be estimated using a formula similar to Eq.~(\ref{eq:fetas2}). A simpler procedure that gives almost identical results (to within $1/4$\%) for the correctly tuned $s$~mass uses
\begin{equation}
	am_s^\mathrm{tuned} \approx am_s 	
	\left(\frac{m_{\eta_s}}{am_{\eta_s}^\mathrm{lat}/a}\right)^2
\end{equation}
where the lattice spacing is obtained from one of the two recipes above (or any other). Again in typical simulations, one guesses a value for $am_s$ and then uses $\eta_s$ results for this mass, together with this formula, to refine the initial guess.

We compare lattice spacings determined using each of our two recipes in Table~\ref{tab:alatt-r1-fetas}. As expected, the lattice spacings are very different on the coarser lattices. This is because $a^2$~errors differ between the $r_1$ and $\eta_s$ measurements. Also as expected (and required), the two recipes converge for smaller lattice spacings, as $a^2$~errors in both types of measurement become negligible. The errors in each case are comparable. We also include values for the correctly tuned $s$-quark mass (in the HISQ formalism) for each configuration set, and for each recipe for the lattice spacing.

In neither of our recipes do we attempt to correct for sea-quark masses that are not correctly tuned. This is standard practice in lattice determinations of the lattice spacing. It pushes any sea-quark mass dependence from $r_1$ or $f_{\eta_s}$ (or whatever is used to determine the lattice spacing) into the other measurements of interest. This is a small effect for $r_1$ and $f_{\eta_s}$, and it is typically extrapolated away together with the sea-quark effects intrinsic to the other measurements.

\begin{table}
\caption{\label{tab:alatt-r1-fetas} Lattice spacings (in fm) and $s$-quark masses (in lattice units) determined using our $r_1$ and $f_{\eta_s}$ recipes. Results are given for each configuration set from Table~\ref{tab:params}. We also list the number of $am_s$ values used in the $\eta_s$ recipe. Note that the estimates converge as the lattice spacings vanish.} 
\begin{center}
\begin{ruledtabular}
\begin{tabular}{cccccc}
Set & \#$am_s$ & $a|_{r_1}$ & $am_s^\mathrm{tuned}|_{r_1}$
	& $a|_{\eta_s}$ & $am_s^\mathrm{tuned}|_{\eta_s}$ \\ \hline
1	&  2 & 0.1456(11) & 0.0613(12) & 0.1583(13) & 0.0724(15) \\
2	&  1 & 0.1465(11) & 0.0622(12) & 0.1595(14) & 0.0736(16) \\ \hline
3	&  1 &  0.1184(9) &  0.0489(9) & 0.1247(10) & 0.0542(11) \\
4	&  3 &  0.1197(9) &  0.0495(9) & 0.1264(11) & 0.0553(11) \\
5	&  2 &  0.1185(9) &  0.0491(9) & 0.1263(11) & 0.0558(12) \\ \hline
6	&  3 &  0.0847(6) &  0.0337(6) &  0.0878(7) &  0.0362(7) \\
7	&  1 &  0.0844(6) &  0.0336(6) &  0.0884(7) &  0.0369(7) \\ \hline
8	&  1 &  0.0592(4) &  0.0226(4) &  0.0601(5) &  0.0233(5) \\ \hline
9	&  2 &  0.0440(3) &  0.0161(3) &  0.0443(4) &  0.0163(3)
\end{tabular}
\end{ruledtabular}
\end{center}
\end{table}

\section{$r_0$}

$r_0/a$ is not determined directly by the MILC Collaboration. Instead
they determine the coefficient of the $1/r$ term in the static potential 
in the region 0.2 --- 0.7 fm. 
If this coefficient is $B$ then:
\begin{equation}
\frac{r_0}{r_1} = \sqrt{\frac{B+C_{r_0}}{B+C_{r_1}}}
\label{eq:r0r1}
\end{equation} 
where $C_{r_0} = 1.65$ and $C_{r_1} = 1.0$~\cite{milc2}. This 
assumes that the same constant $1/r$ coefficient would 
be obtained around $r \approx r_0$ and $r \approx r_1$ and there will be a 
small systematic error, yet to be determined~\cite{milcreview} for this 
assumption. 
$B$ shows dependence on the lattice spacing and the sea quark masses 
as demonstrated in Figure 13 of ~\cite{milcreview}. 
Extrapolating to the continuum and chiral limits gives 
$B= -0.464(7)$, implying from equation~\ref{eq:r0r1}, with the caveats above, that 
$r_0/r_1 = 1.488(5)$. Our value for $r_1$ then gives $r_0 = 0.4661(38)$\,fm. 
This is in agreement with, but more accurate than, the previous 
MILC determination of 0.462(12) fm which used ensembles at fewer 
values of the lattice spacing, but which includes a systematic error of 0.004
from the variation of results with fit range in $r$. Our result also 
agrees with the direct determination from Aoki et al~\cite{fodor} 
of $r_0 = $0.48(1)(1) fm,
which also includes the effect of $u$, $d$ and $s$ sea quarks and comes from an 
analysis with multiple values of the lattice spacing.  

\section{Conclusions}

The accurate determination of the lattice spacing is of critical 
importance to obtaining accurate results from lattice QCD. 
Here we give two ways to do this with sub-1\% errors for the first 
time. 

The first method makes use of the $\approx$ 0.3\% accurate values 
for $r_1/a$ calculated by the MILC collaboration on their ensembles 
(which could also be reproduced on other ensembles with similar 
statistics)
coupled with the 0.8\% accurate value for $r_1$ given here : 
$r_1 = 0.3133(23)$ fm.  
Our result is $1.5\sigma$ from our previous analysis~\cite{alan} using only the $\Upsilon$ 
$2S-1S$ splitting on fewer ensembles and combined with less accurate $r_1/a$ values. 
It is also $1\sigma$ lower than that 
of the MILC collaboration using essentially the same results~\cite{milc2}. It is in agreement with, but 
slightly more accurate than a newer result from MILC~\cite{milcreview} of $r_1 = 0.3108 \binom{+30}{-80}$ fm 
using $f_{\pi}$ data across a similar 
range of lattice spacing values to our $f_{\eta_s}$ analysis but with ASQTAD valence quarks 
rather than HISQ quarks~\cite{milcreview}.  

The second method is possibly simpler (in the absence of $r_1/a$ values) 
since it relies only 
on a standard meson spectrum calculation that would 
automatically be included in many lattice analyses. 
The mass and the decay constant of the 
$\eta_s$ can be determined to better than 0.25\% given 
similar statistics to those we have used here and 
provided a quark formalism is used in which the 
PCAC relation holds so that the decay constant has no 
renormalisation. Then the physical values for $f_{\eta_s}$ 
and $m_{\eta_s}$ that have been determined here can be 
used to find both the tuned value of the strange mass, by 
interpolation in $f_{\eta_s}/m_{\eta_s}$ to 0.2647(18) and 
the lattice spacing, from taking $f_{\eta_s}$=0.1815(10) GeV at the tuned point.

The two methods are compared for the MILC ensembles in Table~\ref{tab:alatt-r1-fetas}.

To improve these methods so that errors below 0.5\% are possible 
will require improvements in the chiral analysis determining the 
$\eta_s$ parameters. These can be gauged from the error budgets in 
Tables~\ref{tab:r1-error} and~\ref{tab:piKetas-errors}. 
Key improvements that are certainly possible are statistical errors in the lattice results 
and accurate lattice data closer to the chiral and 
continuum limits. Improvements to other methods of determining 
the lattice spacing, such as that using the $\Upsilon$ spectrum 
and $m_{D_s}-m_{\eta_c}/2$ discussed here are important for cross-checks 
of systematic effects.

{\bf{Acknowledgements}} We are grateful to the MILC collaboration for the use of 
their configurations and to Doug Toussaint, Steve Gottlieb and Claude Bernard for information 
on $r_1/a$ values. 
We are grateful to other members of the HPQCD collaboration
for useful discussions. 
Computing was done at USQCD's Fermilab cluster, the Ohio Supercomputer Centre 
and the Argonne Leadership 
Computing Facility at Argonne National Laboratory, which is supported by 
the Office of Science of the U.S. Department of Energy under constract DOE-AC02-06CH11357. 
We acknowledge the use of Chroma~\cite{chroma} for part of our analysis. 
This work was supported by 
the Leverhulme Trust,  MICINN, NSF, the Royal Society, the Scottish Universities Physics 
Alliance and STFC. 

\appendix

\section{$f_{\pi}$, $f_{K}$ and $f_{\eta_s}$}
\label{sec:chiral1}
In~\cite{fds}, we analyzed simulation results for pion and kaon masses and decay constants obtained using the HISQ action for the valence quarks, with gluon configurations from MILC, produced using the ASQTAD action for the ($n_f=3$) light sea quarks. We described how to extrapolate these results to the correct light-quark masses and to zero lattice spacing, obtaining decay constants that agree well with experiment. 

\begin{figure}
	\begin{center}
		\includegraphics{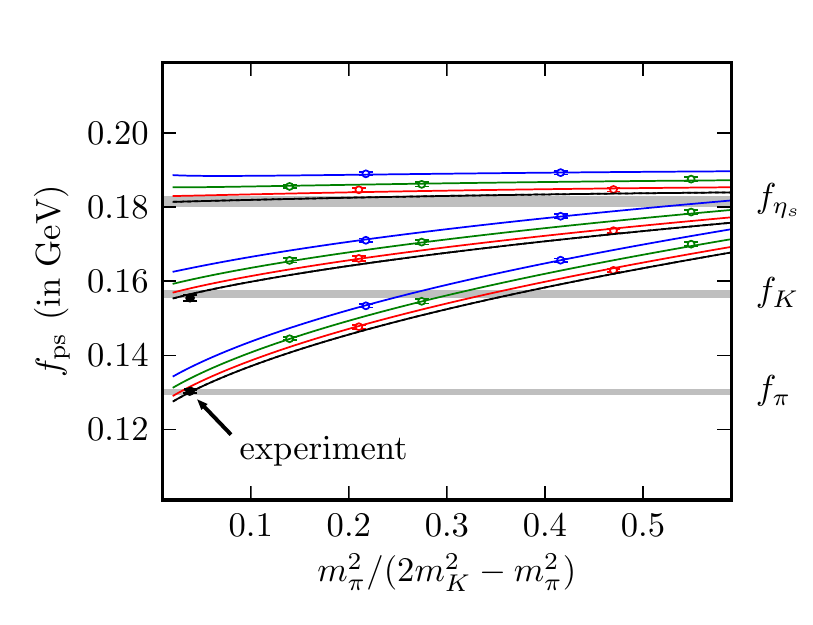}
		\vspace{-3ex}
	\end{center}
	\caption{The pseudoscalar decay constants plotted versus quark mass;
	 $m_\pi^2/(2m_K^2-m_\pi^2)$ is approximately the ratio of the $u/d$ to $s$
	 quark masses: $m_l/m_s$. The fit data is from lattice simulations with
	three different lattice spacings; results decrease with decreasing lattice
	spacing. The data have been adjusted to correspond to points where the
	sea-quark masses correspond to the valence masses. The lines are from 
	the tuned fit function for each of the three lattice spacings. The bottom
	line in each group is the extrapolation to $a=0$. The gray bands indicate
	final values from the fit for the physical decay constants for all three
	mesons; the leftmost data points for $f_\pi$ and $f_K$ are the current
	experimental values.
	}
	\label{fig:fmpi2}
\end{figure}

Here we reuse our earlier simulation results, which are summarized in Table~\ref{tab:piK} (and in Table~\ref{tab:etas} for the $\eta_s$), to extract a value for the static-quark potential parameter~$r_1$. More importantly, we also extract from this analysis continuum values for the mass and decay constant of the $\eta_s$~meson. The masses and decay constants in Table~\ref{tab:piK} are obtained using the procedure described in Section~\ref{sec:fetas} for analyzing $\eta_s$~correlators; we treat all mesons the same way.

% \begingroup
% \squeezetable
\begin{table}
	\caption{\label{tab:piK}
	Simulation results for pseudoscalar meson decay constants and
	 masses (in lattice units) for several different lattice parameter sets 
	(see Table~\ref{tab:params}), $u/d$ valence-quark masses $m_l$, and $s$ 		
	valence-quark masses $m_s$.}
	\begin{ruledtabular}
	\begin{tabular}{cllcccc}
	Set & $am_l$ & $am_s$ 
	& $af_\pi$ & $am_\pi$ & $af_K$ & $am_K$ 
	\\ \hline
	1 & 0.0132 & 0.066 & 0.1152(3) & 0.2408(6) & 0.1290(4) & 0.4081(6) \\
	2 & 0.0264 & 0.066 & 0.1254(4) & 0.3348(6) & 0.1345(4) & 0.4399(7) \\
	\hline
	3 & 0.0067 & 0.0537 & 0.0889(3) & 0.1567(4) & 0.1020(3) & 0.3242(5) \\
	4 & 0.01365 & 0.05465 & 0.0957(4) & 0.2222(5) & 0.1060(4) & 0.3463(6) \\
	5 & 0.0278 & 0.0525 & 0.1041(3) & 0.3113(6) & 0.1095(4) & 0.3727(6) \\
	\hline
	6 & 0.00705 & 0.0366 & 0.0647(2) & 0.1377(4) & 0.0731(3) & 0.2375(4) \\
	7 & 0.01635 & 0.03635 & 0.0710(2) & 0.2050(4) & 0.0759(2) & 0.2594(4)
	\end{tabular}
	\end{ruledtabular}
\end{table}
% \endgroup

To extract a continuum value for $r_1$, we fit the decay constant data for lattice ensemble~$i$ in Table~\ref{tab:piK} (and Table~\ref{tab:etas} for $af_{\eta_s}$) to
\begin{equation}
	a_i f_\mathrm{ps}(\xa,\xb,\xls,\xss,a_i)
\end{equation}
where $f_\mathrm{ps}$ is the formula from Appendix~\ref{sec:chiral2} and $(a,b)$ labels the valence quarks: $(l,l)$ for pions, $(l,s)$ for kaons and $(s,s)$ for $\eta_s$s. The mass parameters $\xa$, $\xb$\,\ldots\,are computed from the simulation masses in Table~\ref{tab:piK}.
Parameter~$r_1$ enters through the lattice spacing, which we take to be
\begin{equation}
	a_i = \frac{r_1}{(r_1/a)_i}
\end{equation}
where values for $(r_1/a)_i$ are given in Table~\ref{tab:params}. We fit data from the pion, kaon and $\eta_s$ simultaneously since all of the fitting parameters are universal.

Our analysis here differs in three ways from our previous paper~\cite{fds}.
First we are including the $\eta_s$ in our simultaneous analysis of the different pseudoscalar mesons; before we only included $\pi$~and $K$~mesons. Second we have re-expressed chiral perturbation theory in terms of pion and kaon masses rather than quark masses. This simplifies the analysis and also gives more reliable estimates for infrared quantities like chiral logarithms. We take the pion and kaon masses corresponding to the sea quark masses from~\cite{milc2} for ensemble sets (3,4,6,7). Results for the other ensembles are not published so we generate approximate meson masses to go with the sea quarks by multiplying the meson masses for the valence quarks (Table~\ref{tab:piK}) by $(m^\mathrm{sea}/m^\mathrm{val})^{1/2}$ (after converting HISQ quark masses into ASQTAD quark masses using Eq.~(\ref{hisq/asq})). Replacing quark masses with meson masses in the chiral formulas gives results that agree well with our previous results.

The third difference from our earlier analysis is that here we require the fitting function to also fit experimental results for $f_\pi$ and $f_K$ at zero lattice spacing. We do this by treating the physical results as additional data to be fit, together with the simulation results, to a single parameterization. In our previous study we fit only simulation results, showing that these agreed with experimental data. Here our goal is different, as we seek an accurate value for~$r_1$. That value is the one that allows the same chiral formulas to fit both our lattice results and the experimental results; $r_1$ is determined, in effect, from the experimental values for~$f_\pi$ and~$f_K$.

Our simulations omit both electromagnetic corrections and isospin-breaking effects. Following~\cite{milc3}, we remove leading-order errors of both sorts by using
\begin{align}
	m_{\hat\pi}^2 &= m_{\pi^0}^2 \\
	m_{\hat K}^2 &= \mbox{$\frac{1}{2}$}\left(m_{K^0}^2 + m_{K^+}^2 - (1+\Delta_E)(m_{\pi^+}^2-m_{\pi^0}^2) \right)
\end{align}
for the physical masses of the pion and kaon. $\Delta_E$ parameterises the violation of Dashen's Theorem which, 
in the chiral limit, states that the $K^+$ and $\pi^+$ have equal electromagnetic corrections, while the $\pi^0$ and $K^0$ have none. 
We take $\Delta_E = 1(1)$. Electromagnetic corrections are also removed from the standard definition of the decay constants, whose values we take to be~\cite{pdg08}:
\begin{equation}
	f_\pi = 0.1304(5)\,\mathrm{GeV}\quad\quad f_K = 0.1555(9)\,\mathrm{GeV}.
\end{equation}

The fitting parameters that are varied in the fit include all of the parameters that define $f_\mathrm{ps}$ (see Appendix~\ref{sec:chiral2}), as well as $r_1$ itself. As discussed in Appendix~\ref{sec:chiral2}, all parameters have priors in our fits. For $r_1$ we take a very broad prior, $r_1 = 0.315(10)$\,fm, that easily encompasses all current estimates; it has little impact on the final results.

The results of our fit are show in Figure~\ref{fig:fmpi2}. The fit is excellent, with a $\chi^2$ per degree of freedom of~0.4. Our main results are
physical (\emph{i.e.}, continuum) values for $r_1$ and for the decay constant and mass of the $\eta_s$:
\begin{align} \label{eq:chiralresults}
	r_1 &= 0.3190(45)(20)\,\mathrm{fm}, \\ \nonumber
	f_{\eta_s} &= 0.1815(10)(2)\,\mathrm{GeV}, \\ \nonumber
	m_{\eta_s} &= 0.6858(38)(12)\,\mathrm{GeV} \\ \nonumber
	f_{\eta_s}/m_{\eta_s} &= 0.2647(18)(1)
\end{align}
By ``physical'' we mean extrapolated to zero lattice spacing and the correct, physical values for the quark masses. We quote two errors here. The first is the fitting error, representing uncertainties from simulation statistics, and from the chiral and lattice-spacing extrapolation. A detailed breakdown of these errors is given in Table~\ref{tab:piKetas-errors}. The second error is equal to the size of the finite-volume correction. As discussed in Appendix~B, finite-volume corrections are somewhat ambiguous for staggered-quark formalisms like HISQ. We choose to include finite-volume corrections, but, to be conservative, take half the size of the correction as an uncertainty.

\begin{table}
	\caption{\label{tab:piKetas-errors}
	Extrapolation and other errors in our results from the chiral 
	analysis of $\pi$, $K$, and $\eta_s$ masses and decay constants. 
	Finite-volume errors are dealt with separately (see text).}
	\begin{ruledtabular}
	\begin{tabular}{rllll}
		& $r_1$ & $f_{\eta_s}$ & $m_{\eta_s}$ & $f_{\eta_s}/m_{\eta_s}$ \\
		\hline
	           $a^2$ extrapolation &  0.6\% &  0.2\% &  0.3\% &  0.3\%\\
	          $m_q$ extrapolations &  0.7 &  0.2 &  0.2 &  0.2\\
	           $r_1/a$ uncertainty &  0.2 &  0.1 &  0.1 &  0.1\\
	  initial uncertainty in $r_1$ &  0.6 &  0.1 &  0.0 &  0.1\\
	experimental errors in $\pi$, $K$ &  0.2 &  0.2 &  0.2 &  0.3\\
	            statistical errors &  0.7 &  0.3 &  0.4 &  0.5\\
	\hline
	               Total  &  1.4\% &  0.5\% &  0.5\% &  0.7\%

	\end{tabular}
	\end{ruledtabular}
\end{table}

We need the physical $\eta_s$ results for our analysis in Section~\ref{sec:fetas} of the $\eta_s$ decay constant~$f_{\eta_s}$. These fit results have statistical correlations with each other, as well as with the output value of $r_1$, the values of $r_1/a$ used in the fit, and the simulation results for $af_{\eta_s}$ (Table~\ref{tab:etas}). We used the fit here to compute means and a covariance matrix for all of these quantities, and this is used as input data in the $f_{\eta_s}$ analysis of Section~\ref{sec:fetas}.

Note that the values for the $\eta_s$ mass and decay constant agree to better than a percent with the leading-order expectations from chiral perturbation theory: $(2m_K^2-m_\pi^2)^{1/2}$ and $2f_K-f_\pi$, respectively. Our analysis above, however, goes far beyond leading order (see Appendix~\ref{sec:chiral2}).
Our $\eta_s$ results are also quite independent of the input 
prior for $r_1$; taking 0.3133(23)\,fm as the prior, for example, causes shifts 
that are smaller than a quarter of a standard deviation. The 
$\eta_s$ parameters are most sensitive to the physical parameters
for the pion and kaon. They can easily be corrected should there 
be small shifts in the values derived from experiment for $f_{\pi}$ and $f_K$.
The changes in the $\eta_s$ parameters would be: 
\begin{align}
\Delta f_{\eta_s} &= 0.6 \Delta f_K + 0.2 \Delta f_{\pi} \\
\Delta m_{\eta_s} &= 0.2 \Delta f_K - 0.6 \Delta f_{\pi}, 
\end{align}
where $\Delta f_{\pi}$ and $\Delta f_K$ are changes in the pion and 
kaon decay constants from the values used here. 

\section{ Augmented Chiral Formulas}
\label{sec:chiral2}
We model light-quark pseudoscalar masses and decay constants 
using partially-quenched chiral perturbation theory, augmented with
corrections for the finite lattice spacing. For simplicity we re-express 
chiral perturbation theory in terms of pion and kaon masses, in place of the 
quark masses, using
\begin{align}
	\xl &= \frac{m_\pi^2/2}{\Lambda_\chi^2} \approx 0.007 \\
	\xs &= \frac{m_K^2-m_\pi^2/2}{\Lambda_\chi^2} \approx 0.17
\end{align}
as expansion parameters, where
\begin{equation}
	\Lambda_\chi \equiv 4\pi f_\pi/\sqrt{2}
	\approx 1.2\,\mathrm{GeV}.
\end{equation}
We use the formulas through next-to-leading order from~\cite{sharpe-pqX},
together with higher-order corrections in $x_l$ and $x_s$ and finite-$a$ 
corrections. For example, we model the mass and 
lattice spacing dependence of the decay constants using
\begin{equation}
	f_\mathrm{ps}(\xa,\xb,\xls,\xss,a) = f^\mathrm{NLO}
	+ \delta f_\chi + \delta f_\mathrm{lat}
\end{equation}
where $f^\mathrm{NLO}$ is the chiral formula through next-to-leading order, $\delta f_\chi$ is the continuum correction due to higher-order mass corrections, $\delta f_\mathrm{lat}$ is the correction due to the finite lattice spacing, and $(a,b)$ labels the valence quarks: $(l,l)$ for pions, $(l,s)$ for kaons, and $(s,s)$ for~$\eta_s$s.

\begin{figure}
	\begin{center}
		\vspace*{3ex}
		\includegraphics{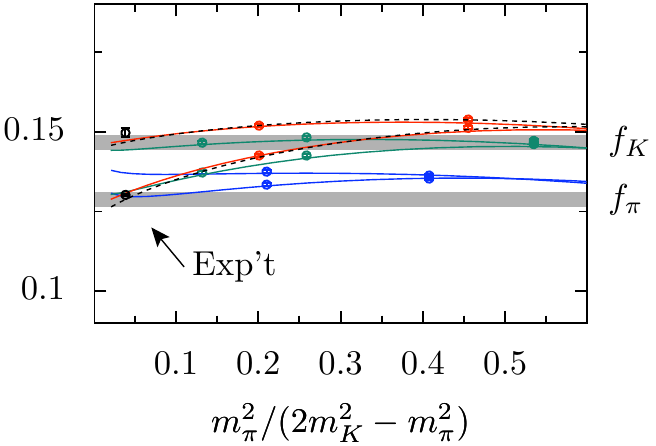}\\[2ex]
		\includegraphics{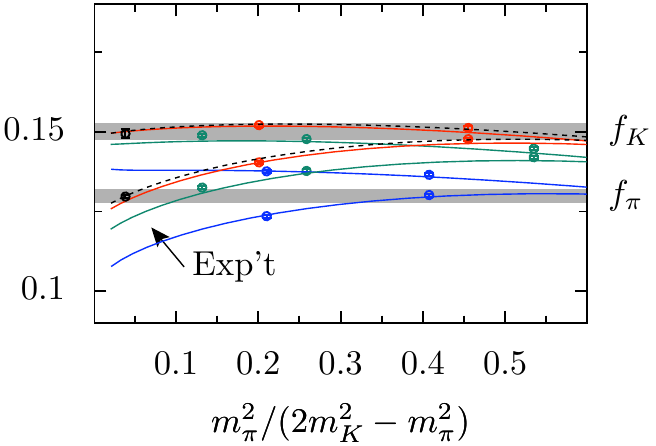}
		\vspace{-2ex}
	\end{center}
	\caption{Fits to two different sets of fake data for pion and kaon decay constants with very different $a^2$ behavior from each other and from the real simulation data (Figure~\ref{fig:fmpi2}). The ``experimental'' points indicated in each case correspond to the exact results, extracted from the formulas used to generate the fake data.}
\label{fig:fakedata}
\end{figure}

Our simulation results are not sufficiently accurate to resolve the difference between high-order polynomials in $x_l$ and $x_s$ and high-order logarithms, so we keep just the polynomials:
\begin{align}
 \delta f_\chi \equiv f_0 & \left( c_1 (\xa+\xb)^2 + c_2 (\xa-\xb)^2 \right.
	\\ \nonumber
	&+ c_3 (\xa+\xb)(2\xls+\xss) 
	\\ \nonumber
	&+ c_4 (2\xls+\xss)^2 + c_5 (2(\xls)^2+(\xss)^2)   
	\\ \nonumber
	& \left.+ c_6 (\xa+\xb)^3 + c_7 (\xa+\xb)(\xa-\xb)^2 \right),
\end{align}
where $f_0$ is the bare decay constant in chiral perturbation theory and the 
$c_i$ are expected to be~${\cal O}(1)$, except for sea-quark terms where the 
coefficients should be 3--5~times smaller.
Still higher-order terms are smaller than 0.1\% and so negligible, as are the last few terms in practice.

Following~\cite{fds}, we model the $a^2$ dependence using a mixture of terms that depend upon $a^2$ and $x_l$:
\begin{align} \label{a2-corrn}
	\delta f_\mathrm{lat} = f_0 &\left( d_1 (a\lamqcd)^2  \alpha_s 
		+ d_2  (a\lamqcd)^2 \alpha_s^3 \right.
	\\ \nonumber
	& + d_3 (a\lamqcd)^2\log(x_l)  \alpha_s^3 
	\\ \nonumber
	& \left.+ d_4 (a\lamqcd)^4 + d_5(a\lamqcd)^5 \right),
\end{align}
where we set $\alpha_s = \alpha_V(2/a)$, $\lamqcd^2=(4m_K^2-m_\pi^2)/3$ as in~\cite{sharpe-pqX}. ($\lamqcd$ is also the ultraviolet scale in the chiral logarithms.). We allow the coefficients to have mass dependence
\begin{align}
	d_i &= d_{i1} + d_{i2}(\xa+\xb) 
	\\ \nonumber
	&+ d_{i3} (2\xls+\xss) + d_{i4} (\xa^2+\xb^2).
\end{align}
Again the $d_{ij}$ are expected to be~${\cal O}(1)$, 
except for terms involving sea-quark 
terms which should be 3--5~times smaller.
The highest-order terms in these expansions are already negligible, making further terms irrelevant. We include the $\log(x_l)$ term in Eq.~(\ref{a2-corrn}) to allow for non-analytic behavior at small~$x_l$, although in practice it is negligible
in our fits.

We included priors in our fitting analysis for each of the parameters in $f^\mathrm{NLO}$ and for all the $c_i$s and $d_{ij}$s. These are initial estimates for each parameter that function as extra ``data'' and allow us to account (in our error estimates) for the uncertainties in these parameters, even when they are largely unconstrained by our simulation data. The parameters in $f^\mathrm{NLO}$ are well determined by our data; we use very broad priors for these, which have no impact on the final errors. We use a prior of $0(1)$ for each of the $c_i$s and $d_{ij}$s, except for terms involving sea-quark masses in which case we use~$0.0(3)$. 

As reported in~\cite{fds}, we have tested these fitting formulas extensively by using formulas from partially-quenched staggered chiral perturbation theory, with randomly selected coefficients and randomly generated higher-order corrections in the masses and $a^2$, to generate fake data sets for the same masses and lattice spacings used in our analysis here. We added statistical noise to the fake data that was comparable in magnitude to that in our real simulation data, with similar correlations. We then fit the fake data using the formulas above, together with the Empirical Bayes method~\cite{bayes} to set a prior for the expansion parameters ($c_i$ and $d_{ij}$). In each case we could compare extrapolated results from our analysis of the fake data with the exact results, since we knew the underlying formula used to generate the fake data. We ran tests for several hundred cases. As expected, we found that 70\% of the time the extrapolated results were within one standard deviation of the exact results. Two examples, shown in Fig.~\ref{fig:fakedata}, illustrate how effective our formulas are in handling $a^2$~dependence that is much larger and much more complex than we see in our actual simulation results (Figure~\ref{fig:fmpi2}),

The logarithms in the NLO chiral formulas reflect infrared sensitivity. These terms are sensitive to the finite volume of our lattice at the level of~0.1--1\% for the decay constants (less for masses). We add finite-volume corrections to the logarithms which we obtain by recomputing the one-loop chiral corrections that lead to logarithms using finite-volume sums instead of integrals in momentum-space, and subtracting them from the infinite volume results. These corrections are quite sensitive to the meson mass, which raises an issue since in staggered-quark formalisms like HISQ each pseudoscalar meson comes in several different ``tastes'', all of them heavier than the Goldstone meson whose mass we use in our formulas. Taste splittings are $a^2$ corrections, which vanish in the continuum limit, and most of the effects of these we model with our corrections~Eq.~(\ref{a2-corrn}) (which we have tested, as discussed in the previous paragraph). The finite-volume corrections, however, are particularly sensitive to meson masses, so we use an ``effective'' pseudoscalar mass when we calculate them:
\begin{equation}
	(m_{ab}^\mathrm{eff})^2 = (m_{ab}^\mathrm{gs})^2 + g_m (a/r_1)^2
\end{equation} 
where $m_{ab}^\mathrm{gs}$ is the Goldstone meson's mass. We expect 
$g_m\approx 0.2\,\mathrm{GeV}^2$. We allow $g_m$ to float in our fits, treating it as a fit parameter. We use $0.2(6)$ as our prior. Our fit favors a nonzero value for $g_m$, giving~$g_m=0.2(3)$ which is consistent with expectations.

\section{Sea-Quark Masses}
\label{sec:sea-quark-masses}
We include terms in each of our fitting functions that correct for the discrepancies $\delta m^\mathrm{sea}_q$ between the bare sea-quark masses used in the simulation and the physically correct bare quark masses (that is, the ones that give correct masses for the $\pi$, $K$, and $\eta_s$). Our estimates for the correct $s$-quark masses (in lattice units) for each ensemble are given in Table~\ref{tab:alatt-r1-fetas}; the $u/d$ mass is 27.8(3) times smaller~\cite{fds}. These masses, however, are for HISQ quarks, while the sea quarks were all analyzed using the ASQTAD formalism. Quark masses in the two formalisms can be related to each other, ensemble by ensemble, by comparing $\pi$ and $\eta_s$ masses for mesons whose valence quarks are either HISQ or ASQTAD quarks. HISQ masses and ASQTAD masses are equivalent when they give the same $\pi$ and $\eta_s$ masses. The ratio of a HISQ mass to the corresponding ASQTAD mass determined in this way should be almost independent of the valence-quark mass, but will depend somewhat on the lattice spacing and weakly on the sea-quark masses. We have compared ASQTAD data from~\cite{milc2} for ensembles~3,4,6,7 with our results in Table~\ref{tab:piK} to obtain the following simplified parameterization for the ratio of HISQ to ASQTAD quark masses:
\begin{equation}
	\frac{am^\mathrm{hisq}}{am^\mathrm{asq}} = 1.158\,
	\frac{1+0.44\, (a/r_1)^2}{1+0.009\, (am_\mathrm{tot}^\mathrm{asq}/am_{s}^\mathrm{tuned})}
	\label{hisq/asq}
\end{equation}
where $m_s^\mathrm{tuned}$ is the tuned HISQ mass given in Table~\ref{tab:alatt-r1-fetas} and $m_\mathrm{tot}^\mathrm{asq}$ is the sum of the three sea-quark ASQTAD masses for that ensemble. This formula is accurate to a few percent.


\begin{thebibliography}{99}
\bibitem{fnpdg08} Taking the orbital excitation energy as the spin-averaged mass of 
$\chi$ (1P) states minus the spin-average of vector and pseudoscalar (1S) states 
gives 457\,MeV from experiment for both $b\overline{b}$ and $c\overline{c}$. For the spin-average of 
2S minus 1S states $c\overline{c}$ gives 606(1)\,MeV and for $b\overline{b}$,  
$(581-h/4)$\,MeV, where $h$ is
the unknown hyperfine splitting at the 2S level (of order 30-40\,MeV)~\cite{pdg08}.
\bibitem{corrn} Note that there is a typographical error in the lattice Hamiltonian given in~\cite{alan}. 
The Hamiltonian used there and in this work are identical.
\bibitem{ps} Smearings suitable for ground-state hadrons are useful where there is a problem 
with expoonentially falling signal/noise, as in the $B$ system. This is not an issue for 
the $b\overline{b}$ system.
\bibitem{fnsyst} The impact on our final value of $r_1$ from our 1.5\% systematic error in $m_{D_s} - m_{\eta_c}/2$ is 
reduced to 1.1\% by the prior for the physical value of $r_1$ ($=0.315(10)$\,fm).
\end{thebibliography}
\end{document}